\documentclass[graybox]{svmult}

\usepackage{helvet}         
\usepackage{courier}        
\usepackage{type1cm}        
\usepackage{makeidx}         
\usepackage{graphicx}        
\usepackage{multicol}        
\usepackage[bottom]{footmisc}

\newcommand{\be}{\begin{equation}}
\newcommand{\ee}{\end{equation}}
\newcommand{\bea}{\begin{eqnarray}}
\newcommand{\eea}{\end{eqnarray}}
\newcommand{\nn}{\nonumber}
\newcommand{\beas}{\begin{eqnarray*}}
\newcommand{\eeas}{\end{eqnarray*}}
\newcommand{\ket}[1]{\left| #1 \right\rangle}
\newcommand{\bra}[1]{\left\langle #1 \right|}

\newcommand{\Tr}{{\rm Tr}}

\newcommand{\vev}[1]{\left\langle #1 \right\rangle}
\newcommand{\binomi}[2]{\left(\begin{array}{c} #1 \\ #2 \end{array}\right)}

\begin{document}

\title*{Many-body localization in large-$N$ conformal mechanics}
\author{Fumihiko Sugino and Pramod Padmanabhan}
\institute{Fumihiko Sugino \at Fields, Gravity \& Strings Group, Center for Theoretical Physics of the Universe, Institute for Basic Science (IBS), Seoul 08826, Republic of Korea \email{fusugino@gmail.com} \and 
Pramod Padmanabhan \at Fields, Gravity \& Strings Group, Center for Theoretical Physics of the Universe, Institute for Basic Science (IBS), Seoul 08826, Republic of Korea \email{pramod23phys@gmail.com}}
\maketitle

\abstract{In quantum statistical mechanics, 
closed many-body systems that do not exhibit thermalization after an arbitrarily long time 
in spite of the presence of interactions are called as many-body localized systems, and recently have been vigorously investigated. 
After a brief review of this topic, we consider a many-body interacting 
quantum system in one dimension, which has conformal symmetry and integrability. 
We exactly solve the system and discuss its thermal or non-thermal behavior. 
}

\section{Introduction}
\label{sec:introduction}
In quantum statistical physics, it is still a big challenge to formulate and understand how systems out of thermal equilibrium settle down to systems in thermal equilibrium, 
although innumerable attempts has been done toward its understanding for over a century. 
Recently, by investigating closed quantum many-body systems and their time evolution for a sufficiently long time, two qualitaitvely different phases have been found in the thermodynamic limit, 
which are referred to as {\em thermalization}/{\em delocalization} and {\em localization}. 
First, we start with a brief review of these phases~\footnote{For review articles, see~\cite{huse,altman,imbre} for example.}.

\subsection{Thermalization}
Let us consider a closed quantum system $S$, for which the Hamiltonian $H$ is defined. The density matrix of the system $\rho$ evolves with the time $t$ as  
\be
\rho(t)= e^{-iHt}\rho(0) \,e^{iHt}. 
\label{rhot}
\ee
Suppose the same system is put in thermal equilibrium at temperature $\beta^{-1}$. Its density matrix is expressed as 
\be
\rho^{({\rm eq})}(\beta) = \frac{1}{Z(\beta)}e^{-\beta H} \qquad \mbox{with} \qquad Z(\beta)=\Tr \,e^{-\beta H}. 
\label{rhoeq}
\ee

%
\begin{figure}[h]
\sidecaption[t]
\includegraphics[height=4.5cm, width=6cm, clip]{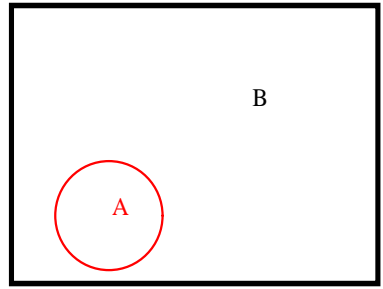}
\caption{The closed system $S$ is the inside of the box. The subregion $A$ is a region bounded by the red circle, and $B=S-A$ is the rest. } 
\label{fig:AB}
\end{figure}
%
Next, we pick any small subregion $A$ in $S$ in real space, and regard $B=S-A$ as a reservior (see Fig.~\ref{fig:AB}). 
The reduced density matrix of $A$ for (\ref{rhot}) and (\ref{rhoeq}) is obtained from $\rho$ by 
tracing out the states belonging to the Hilbert space of the subsystem $B$:  
\be
\rho_A(t)= \Tr_B\, \rho(t), 
\label{rhoA}
\ee
and 
\be
\rho^{({\rm eq})}_A(\beta)= \Tr_B\, \rho^{({\rm eq})}(\beta), 
\ee
respectively. 
Then, we define thermalization as follows. 
\begin{definition}
If
\be
\rho_A(t) \to \rho^{({\rm eq})}_A(\beta)
\ee
as sending $t$ and the volume of $S$ to infinity with the volume of $A$ being fixed, and if it holds for any choice of the subsystem $A$, 
the system $S$ thermalizes for the temperature $\beta^{-1}$. 
\label{def:thermal}
\end{definition}
Note that since in a closed system the density matrix of the total system $\rho(t)$ undergoes unitary time-evolution, $\rho(t)$ does not 
evolve to $\rho^{(\rm eq)}(\beta)$ in general. This brings us to the Eigenstate Thermalization Hypothesis.   

\subsection{Eigenstate Thermalization Hypothesis}
Suppose the initial state $\rho(0)$ is a pure state for an energy eigenstate of the energy $E_n$:
\be
\rho(0)= \ket{E_n}\bra{E_n} \qquad \mbox{with} \qquad H\ket{E_n}=E_n\ket{E_n}. 
\ee
Then, $\rho$ is time-independent: $\rho(t)=\rho(0)$, which leads to $\rho_A(t)=\rho_A(0)$ for any $A$ from (\ref{rhoA}). 
In this case, noting Definition~\ref{def:thermal}, we could expect that all the energy eigenstates are thermalized, which is called as 
the Eigenstate Thermalization Hypothesis (ETH)~\cite{deutsch,srednicki,tasaki,rigol}.   

If ETH holds, the temperature at the thermal equilibrium, denoted by $\beta_n^{-1}$, is determined by 
\be
E_n=\vev{H}_{\beta_n}\equiv \frac{1}{Z(\beta_n)}\,\Tr\left(H\, e^{-\beta_nH}\right).
\ee
The entanglement entropy of the subsystem $A$:
\be
S_A=-\Tr_A\left(\rho_A\ln \rho_A\right)
\ee
coincides with the equilibrium thermal entropy of $A$. In particular, $S_A$ is an extensive quantity, proportional to the volume of $A$. 

However, ETH is a hypothesis, and not true for one class of systems.  
Such systems are called as localized systems. 
 
\subsection{Localized Systems}
A simple example of single-particle localization is given by the one-dimensional Hamiltonian:
\be
H=-\frac{1}{2m} \frac{\partial^2}{\partial x^2} + V_p(x) +V_q(x),
\ee
where $V_p(x)$ is a periodic potential, and $V_q(x)$ is a random noise. 
If the noise is absent ($V_q(x)=0$), the wave function of the particle is oscillating due to the Bloch wave, and delocalized. 
However, when the noise is turned on, the wave function becomes localized as
\be
\psi(x)\sim e^{-\mu_q x} \qquad \mbox{as} \qquad |x|\to \infty
\ee
with a strictly positive constant $\mu_q$. This phenomenon is well-known as the Anderson localization~\cite{anderson,brandenberger}  

Next, we turn to many-body localization (MBL), which takes place in the presence of many-body interactions and for highly excited states. 
A typical example is given by a one-dimensional quantum spin-$1/2$ chain, whose Hamiltonian takes the form 
\be
H=\sum_ih_i\sigma^z_i + J\sum_{<i,\,j>}\vec{\sigma}_i\cdot\vec{\sigma}_j.
\label{H_MBL}
\ee
Here, $i, j\in\{1, 2,\cdots\}$ denote the sites of the system, $h_i$ are random magnetic fields at the site $i$ distributed over the range $[-W, W]$, and 
the second term represents the nearest neighbor interactions of the Pauli spins. 

At  $J=0$, the eigenstates of (\ref{H_MBL}) are product states of the $\sigma^z$ eigenstates: 
$\ket{\sigma^z_1}\otimes\ket{\sigma^z_2}\otimes\cdots$ with $\ket{\sigma^z}=\ket{\uparrow}$ or $\ket{\downarrow}$. 
Each spin variable is completely decoupled and undergoes independent time evolution. 
This system is fully localized, and essentially the same as the above single-particle localization.   
There are strictly local integrals of motions (LIOM) $\sigma^z_i$ ($i=1,2\cdots$), whose supports are on single sites. 

When turning on $J$ but $J\ll W$, the localization property somehow remains. This case is called as MBL. 
There are also LIOM, but they satisfy milder locality condition with exponentially decaying tails in large distances 
(called as quasi LIOM). Such quasi LIOM are constructed, and DC spin transport and energy transport are shown to be absent 
perturbatively and nonperturbaively with respect to the coupling $J$~\cite{basko,imbre0}. 
    
On increasing $J$, the localization ceases and ETH starts to hold eventually. 
Interestingly, there will be a transition between MBL (localized) and delocalized phases around $J\sim W$, which is a new type of phase transition 
between thermal equilibrium and out-of-equilibrium.   
It is expected that the localization is an intriguing phenomenon that protects the system from thermal decoherence and can be useful  
to construct devices for quantum computations. 

However, analyses for MBL have been performed mainly for quantum spin systems. Extension to other quantum systems should be 
important to find new aspects and understand universal properties for localizations. In the rest of this contribution, we construct an integrable model of 
many-body conformal quantum mechanics by using its coalgebra structure, and analyze its thermal or localization properties. 

\section{Many-body interacting model by using coproducts}
\label{sec:hamiltonian}
The conformal group in one dimension, $SL(2,{\bf R})$, is generated by the Lie algebra generators $L_0$, $L_\pm$ satisfying 
\be
[L_0,\, L_\pm]=\pm L_\pm, \qquad [L_+,\, L_-]=-2L_0
\label{SL2R}
\ee
with the quadrartic Casimir 
\be
C=L_0^2-L_0-L_+L_-.
\label{Casimir}
\ee
This is realized in one-dimensional quantum mechanical system~\cite{dealfano} as 
\bea
L_0 & = & \frac14\left(p^2+\frac{g}{x^2}+x^2\right), 
\label{L0}
\\
L_\pm & = &  \frac14\left(-p^2-\frac{g}{x^2}+x^2\right) \mp i\frac14(xp+px) 
\label{Lpm}
\eea
with $[x,\, p]=i$ and $C=-\frac{3}{16}+\frac14g$. 
$L_0$ plays a role of the Hamiltonian. 
For simplicity, we will consider the case of $g=0$, in which the system reduces to a harmonic oscillator. 

\subsection{Coproducts}
In treating $N$-body systems, it is convenient to introduce coproducts denoted by $\Delta^{(k)}$ ($k=2, 3,\cdots,N$). 
Let $L_{a,\,i}$ ($a=0,\,\pm$) be the $L_a$-operator for particle $i$ (or at site $i$). 
$\Delta^{(2)}(L_a)$ acts on two-particle states, which is defined by 
\be
\Delta^{(2)}(L_a) = L_a\otimes 1 + 1\otimes L_a = L_{a,\,1}+L_{a,\,2}. 
\ee
Also, $\Delta^{(2)}(1)=1\otimes 1.$
Then, $\Delta^{(3)}(L_a)$ acting on three-particle states is given as
\bea
\Delta^{(3)}(L_a) & = & ({\bf 1}\otimes \Delta^{(2)})\circ \Delta^{(2)}(L_a) \nn \\
& = & ({\bf 1}\otimes \Delta^{(2)})\circ (L_a\otimes 1 + 1\otimes L_a)  \nn \\
& = & L_a\otimes \Delta^{(2)}(1) + 1\otimes\Delta^{(2)}(L_a) \nn \\
& = & L_a\otimes 1\otimes 1 + 1\otimes (L_a\otimes 1 + 1\otimes L_a) \nn \\
& = & L_{a,\,1} + L_{a,\,2} + L_{a,\,3}, 
\eea
In general, $\Delta^{(k)}(L_a)$ is inductively given as 
\bea
\Delta^{(k)}(L_a) &= & (\overbrace{{\bf 1}\otimes \cdots\otimes{\bf 1}}^{k-2}\otimes\Delta^{(2)})\circ \Delta^{(k-1)}(L_a) \nn \\
& = & L_{a,\,1}+\cdots +L_{a,\,k}. 
\eea
Note that the coproducts act as homomorphism and preserve the algebra~(\ref{SL2R}):
\bea
& & [\Delta^{(k)}(L_0),\,\Delta^{(k)}(L_\pm)] = \pm \Delta^{(k)}(L_\pm), \\
 & & [\Delta^{(k)}(L_+),\,\Delta^{(k)}(L_-)] = -2\Delta^{(k)}(L_0)
\eea
with the quadratic Casimir
\be
\Delta^{(k)}(C)= \left(\Delta^{(k)}(L_0)\right)^2 - \Delta^{(k)}(L_0) 
-\Delta^{(k)}(L_+)\Delta^{(k)}(L_-).  
\ee
We can see that $\Delta^{(k')}(C)$ commutes with $\Delta^{(k)}(L_a)$ for $k'\leq k$. 

\subsection{Hamiltonian}
\begin{figure}[h]
\centering
\includegraphics[height=3cm, width=7cm, clip]{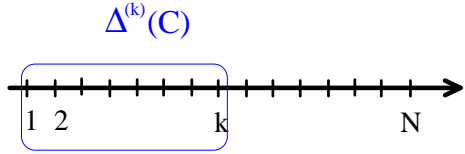}
\caption{The operator $\Delta^{(k)}(C)$ has support on sites $\{1,\,2,\, \cdots, \,k\}$. } 
\label{fig:interaction}
\end{figure}

We consider the Hamiltonian for $N$-particle interacting conformal system as 
\be
H_N=\Delta^{(N)}(L_0) +\sum_{k=2}^N\alpha_k\Delta^{(k)}(C),
\label{HN}
\ee
where the first term describes $N$ free harmonic oscillators, and the rest are interactions with coupling constants $\alpha_k$. 
$\Delta^{(k)}(C)$ is an interaction with support on sites 1 to $k$ as depicted in Fig.~\ref{fig:interaction}. 
The construction of (\ref{HN}) is based on the idea in~\cite{ballesteros,musso}.  
Eventually, we send $N$ to infinity.  

%
%

Since $\Delta^{(N)}(L_0)$ and $\Delta^{(k)}(C)$ ($k=2,\cdots, N$) mutually commute, they give $N$ conserved quantities. 
This implies that the system is integrable. 
However, they are not local in general, and it is nontrivial whether the system exhibits MBL. 
If we choose the coupling constants behaving as 
\be
\alpha_k\sim e^{-k/\xi} \qquad \mbox{with $\xi$ some positive number}, 
\label{alpha_expdecay}
\ee
all the interactions become quasi local and 
the above conserved quantities can be regarded as quasi LIOM. 

In terms of the position and momentum variables, (\ref{HN}) is expressed as 
\be
H_N= \sum_{i=1}^N\frac14\left(p_i^2+x_i^2\right) 
+\sum_{k=2}^N\alpha_k
\left\{\frac14\sum_{1\leq i<j\leq k}M_{ij}^2+\frac{k(k-4)}{16}\right\}
\ee
with $M_{ij}\equiv x_ip_j-x_jp_i$ being an analog of angular momentum operators.

\section{Eigenstates and eigenvalues}
\label{sec:eigen}
In order to exactly solve the system (\ref{HN}), we first consider the lowest weight states (level 0 states) satisfying 
\be
L_{-,\,i}\ket{s}_N=0 \qquad \mbox{for} \qquad i=1,\cdots, N.
\ee
Here, the subscript `$N$' in the state vector is used to denote the $N$-particle state.  
The conditions are solved as 
\be
\ket{s}_N=\ket{r^{(1)}_0,\,\cdots, r^{(N)}_0} \equiv \ket{r^{(1)}_0}\otimes \cdots\otimes \ket{r^{(N)}_0}
\ee
with $\ket{r^{(i)}_0}$ being the eigenstate of $L_{0,\, i}$ with the weight $1/4$ or $3/4$:  
\be
L_{0,\,i}\ket{r^{(i)}_0}=r^{(i)}_0\ket{r^{(i)}_0} \qquad \left(r^{(i)}_0=\frac14, \,\frac34\right).
\ee
The weights $1/4$ and $3/4$ correspond to the ground state energy and the first excited energy of the harmonic oscillator, 
respectively.   
The energy eigenvalue is given by
\be
E_0=R_N + \sum_{k=2}^N\alpha_kR_k\left(R_k-1\right), \qquad R_k\equiv r^{(1)}_0+\cdots +r^{(k)}_0
\label{E0}
\ee

Any state vector in the Fock space can be obtained by successively acting $L_{+,\,i}$ operators on the level 0 states. From the $SL(2,\,{\bf R})$ algebra (\ref{SL2R}), 
the states containing $n$ $L_{+,\,i}$ operators increase the weight by $n$, and correspond to $2n$-th excited states of the harmonic oscillator. 
The Fock space is decomposed as 
\be
{\cal F} = \bigoplus_{r_0^{(1)},\cdots,r_0^{(N)}}{\cal F}_{(r_0^{(1)},\cdots,r_0^{(N)})}
\ee
with 
\be
{\cal F}_{(r_0^{(1)},\cdots,r_0^{(N)})}\equiv \left\{L_{+,\,1}^{k_1}\cdots L_{+,\,N}^{k_N}\ket{s}_N\,;\, k_1, \cdots, k_N=0, 1, 2, \cdots\right\}. 
\ee
$L_{+,\,1}^{k_1}\cdots L_{+,\,N}^{k_N}\ket{s}_N$ is the eigenstate of $\Delta^{(N)}(L_0)$ with the eigenvalue $k_1+\cdots +k_n+R_N$, and 
called as a level $k_1+\cdots +k_n$ state. 

\subsection{Level 1 states} 
We find the following $N$ states of level 1:
\bea 
\ket{v_1}_N & = & \Delta^{(N)}(L_+)\ket{s}_N, \nn \\
\ket{v_{1,\,(1,1)}}_N & = & F_1\left(L_{+,\,1},\,L_{+,\,2}\right)\ket{s}_N, \nn \\
\ket{v_{1,\,(1,2)}}_N & = & F_1\left(\Delta^{(2)}(L_+),\,L_{+,\,3}\right)\ket{s}_N, \nn \\
& & \vdots  \nn \\
\ket{v_{1,\,(1,\,N-1)}}_N & = & F_1\left(\Delta^{(N-1)}(L_+),\,L_{+,\,N}\right)\ket{s}_N, 
\label{level1}
\eea
where $F_1\left(\Delta^{(n)}(L_+),\,L_{+,\,n+1}\right)$ is a linear function of $\Delta^{(n)}(L_+)$ and $L_{+,\,n+1}$ given by
\be
F_1\left(\Delta^{(n)}(L_+),\,L_{+,\,n+1}\right)=-\frac{r^{(n+1)}_0}{R_n}\Delta^{(n)}(L_+) +L_{+,\,n+1}
\label{F1}
\ee
for $n=1,\cdots, N-1$, and hereafter $\Delta^{(1)}(L_+)$ is regarded as $L_{+,\,1}$. 
Notice that 
\bea
& & \Delta^{(m)}(L_-) \ket{v_{1,\,(1,n)}}_N=0, \nn \\
& & F_1\left(\Delta^{(m)}(L_-),\,L_{-,\,m+1}\right)\ket{v_{1,\,(1,n)}}_N=0
\label{level1_p2}
\eea
hold for $m>n$, 
which leads to the orthogonality of the states (\ref{level1}). 

The energy eigenvalues are obtained as 
\be
E_1=R_N+1+\sum_{k=2}^N\alpha_kR_k(R_k-1)
\ee
for $\ket{v_1}_N$, and 
\be
E_{1,\,(1,n)}=R_N+1+\sum_{k=2}^n\alpha_kR_k(R_k-1) +\sum_{k=n+1}^N\alpha_k(R_k+1)R_k
\ee
for $\ket{v_{1,\,(1,n)}}_N$. 

\subsection{Level $p$ states} 
General level $p$ states are obtained as  
\bea
\ket{v_p}_N & = & \left(\Delta^{(N)}(L_+)\right)^p\ket{s}_N, \nn \\
\ket{v_{p,\,(m_1,n_1),\,\cdots, (m_q,n_q)}}_N & =  & \left(\Delta^{(N)}(L_+)\right)^{p-m_1-\cdots -m_q} \nn \\
 & & \times F_{m_1}\left(\Delta^{(n_1)}(L_+),\,L_{+,\,n_1+1}\right)_{+m_2+\cdots +m_q} \nn \\
 & & \times F_{m_2}\left(\Delta^{(n_2)}(L_+),\,L_{+,\,n_2+1}\right)_{+m_3+\cdots +m_q} \nn \\
 & & \times \cdots \nn \\
 & & \times F_{m_{q-1}}\left(\Delta^{(n_{q-1})}(L_+),\,L_{+,\,n_{q-1}+1}\right)_{+m_q} \nn \\
 & & \times F_{m_q}\left(\Delta^{(n_q)}(L_+),\,L_{+,\,n_q+1}\right)\ket{s}_N,
 \label{levelp}
\eea 
where $q$ runs from 1 to $p$, and $m_1,\cdots m_q\in\{1,\cdots, p\}$ satisfy $\sum_{i=1}^qm_i\leq p$. 
The integers $n_i$ should be taken as $N-1\geq n_1>n_2>\cdots>n_q\geq 1$. 
$F_m\left(\Delta^{(n)}(L_+),\,L_{+,\,n+1}\right)$ is a degree-$m$ homogeneous polynomial of $\Delta^{(n)}(L_+)$ and $\,L_{+,\,n+1}$, 
whose explicit form is  
\bea
F_m\left(\Delta^{(n)}(L_+),\,L_{+,\,n+1}\right) & = & c^{(m)}_0\left(\Delta^{(n)}(L_+)\right)^m
+c^{(m)}_1\left(\Delta^{(n)}(L_+)\right)^{m-1}L_{+,\,n+1}\nn \\
& & \hspace{-1cm} +\cdots + c^{(m)}_{p-1}\Delta^{(n)}(L_+) \left(L_{+,\,n+1}\right)^{m-1}+\left(L_{+,\,n+1}\right)^m 
\label{Fm}
\eea
with the coefficients 
\be
c^{(m)}_k\equiv (-1)^{m-k}\binomi{m}{k}\frac{\Gamma\left(2r^{(n+1)}_0+m\right)}{\Gamma\left(2r^{(n+1)}_0+k\right)}
\frac{\Gamma\left(2R_n\right)}{\Gamma\left(2R_n+m-k\right)}. 
\label{cmk}
\ee
Note that (\ref{Fm}) is independent of the couplings $\alpha_k$'s. 
$F_m\left(\Delta^{(n)}(L_+),\,L_{+,\,n+1}\right)_{+\ell}$ denotes (\ref{Fm}) with every $R_n$ appearing in (\ref{cmk}) replaced by $R_n+\ell$. 
The states in (\ref{levelp}) consist of mutually orthogonal $\binomi{p+N-1}{p}$ states. 
All of the states have no dependence on the couplings, which comes from the Hamiltonian (\ref{HN}) consists of the mutually commuting operators.  

The norms of the states are computed as 
\bea
& & \left|\!\left|\ket{v_p}_N\right|\!\right|^2 = p!\frac{\Gamma\left(2R_N+p\right)}{\Gamma\left(2R_N\right)}, \\
& & \left|\!\left|\ket{v_{p,\,(m_1,n_1),\cdots,(m_q,n_q)}}_N\right|\!\right|^2 =\left(p-M_1\right)!\frac{\Gamma\left(2R_N+M_1+p\right)}{\Gamma\left(2R_N+2M_1\right)} \nn \\
& & \hspace{2cm}\times \prod_{a=1}^q\left[m_a!\frac{\Gamma\left(2r^{(n_a+1)}_0+m_a\right)}{\Gamma\left(2r^{(n_a+1)}_0\right)}\frac{\Gamma\left(2R_{n_a}+2M_{a+1}\right)}{\Gamma\left(2R_{n_a}+2M_{a+1}+m_a\right)}\right.\nn \\
& & \hspace{3cm}\left.\times \frac{\Gamma\left(2R_{n_a+1}+2M_{a}-1\right)}{\Gamma\left(2R_{n_a+1}+2M_{a+1}+m_a-1\right)}\right]
\eea
with 
\be
M_a\equiv\sum_{k=a}^qm_k.
\label{Ma}
\ee
The energy eigenvalues are 
\be
E_p=R_N+p+\sum_{k=2}^N\alpha_kR_k(R_k-1)
\ee
for $\ket{v_p}_N$, and 
\bea
E_{p,\,(m_1,n_1),\cdots,(m_q,n_q)} & = & R_N+p +\sum_{k=2}^{n_q}\alpha_kR_k(R_k-1) \nn\\
& & +\sum_{\ell=2}^q\sum_{k=n_\ell+1}^{n_{\ell-1}}\alpha_k \left(R_k+M_{\ell}\right) \left(R_k+M_{\ell}-1\right) \nn\\
& & +\sum_{k=n_1+1}^N\alpha_k\left(R_k+M_1\right)\left(R_k+M_1-1\right)
\eea
for $\ket{v_{p,\,(m_1,n_1),\,\cdots, (m_q,n_q)}}_N$. 

We can see that all the level $p$ states are degenerate for the free case, while 
the degeneracy is completely resolved by turning on the couplings $\alpha_k$. 
Note for the choice (\ref{alpha_expdecay}), the level splitting between states with different $m_j$'s is of the order 
$O\left(e^{N/\xi}\right)$, which yields continuous spectrum at large $N$. 
This seems a situation in which thermalization takes place. 
On the other hand, there are quasi local LIOM that support MBL as we have seen in section~\ref{sec:hamiltonian}. 
Thus, it is interesting to see which property of ETH and MBL is realized in this case. 

\section{Entanglement entropy}
\label{EE}
Let us start with the density matrix for the pure state:
\be
\rho=\frac{1}{\left|\!\left|\ket{v_{p,\,(m_1,n_1),\cdots,(m_q,n_q)}}_N\right|\!\right|^2} \ket{v_{p,\,(m_1,n_1),\cdots,(m_q,n_q)}}_N\bra{v_{p,\,(m_1,n_1),\cdots,(m_q,n_q)}}. 
\label{rho_pure}
\ee 
We divide the total system $S=\{1,2,\cdots,N\}$ into a small subsystem $A=\{N-\nu+1,\cdots, N\}$ with $\nu\ll N$ and 
the rest $B=\{1,2,\cdots,N-\nu\}$. 
For simplicity, we consider the case of $n_1\leq N-\nu-1$, in which all the $F_m$ operators in (\ref{levelp}) act only on $B$. 
For such pure states, the reduced density matrix $\rho_A$ takes a diagonal form with each diagonal entry taking a simple form:
\be
\lambda_{A,\,\tilde{n}}\equiv \binomi{p-M_1}{\tilde{n}}\frac{B\left(2R_{N-\nu}+2M_1+\tilde{n},\, 2\bar{R}_\nu +p-M_1-\tilde{n}\right)}{B\left(2R_{N-\nu}+2M_1,\, 2\bar{R}_\nu\right)},
\label{lambda}
\ee
where 
\be
\bar{R}_\nu\equiv \sum_{i=N-\nu+1}^Nr^{(i)}_0,
\label{Rbar} 
\ee
and $\tilde{n}$ runs from 0 to $p-M_1$. 

We find the large-$N$ behavior of the entanglement entropy
\be
S_A=-\sum_{\tilde{n}=0}^{p-M_1}\lambda_{A,\,\tilde{n}} \ln\lambda_{A,\,\tilde{n}}
\ee
in the following two cases:
\begin{itemize}
\item 
For $p-M_1\ll R_N+M_1$ (case 1), 
\be
S_A\sim \bar{R}_\nu\frac{p-M_1}{R_N+M_1}\ln\left(R_N+M_1\right).
\label{SA_case1}
\ee
Since $\bar{R}_\nu$ grows with $\nu$ (the volume of $A$), this result exhibits the volume-law like behavior although the multiplicative factor 
$\frac{p-M_1}{R_N+M_1}\ln\left(R_N+M_1\right)$ is tiny for the case. 
\item
For $p-M_1\gg R_N+M_1$ (case 2), 
\be
S_A\sim \ln (p-M_1).
\label{SA_case2}
\ee
This result is independent of $\nu$, and exhibits the area law, which supports the localization phase. 
\end{itemize}

In the case 1, the energy is relatively lower, but the result (\ref{SA_case1}) seems to support thermal like phase. 
On the other hand, in the case 2, the energy is relatively higher, and the result (\ref{SA_case2}) suggests localization.  
Interestingly, because the states (\ref{rho_pure}) do not depend on the couplings $\alpha_k$, the above results hold for any choice of $\alpha_k$. 
In particular, the result means that there are some highly excited states which exhibit the area law behavior (\ref{SA_case2}) even in the presence of nonlocal interactions.  
It is also interesting to analyze the case in which $p-M_1$ is comparable to $R_N+M_1$ (the intermediate region of the cases 1 and 2), 
and to see how the volume-law like behavior changes to the area law.   

\section{Discussion}
\label{sec:discussion}
In this contribution, first we have briefly reviewed topics on quantum thermalization and localization. 
Second, we have constructed an integrable model with many-body interactions 
by using coproducts, and obtained the exact spectrum of the model. 
Third, by computing the entanglement entropy, we have found a localization property 
in highly excited states in spite of nonlocal interactions. 
We guess that this captures a new aspect of localization, which has not been seen yet.  

Since the entanglement entropy does not depend on the couplings, it will be interesting to analyze other quantities that are sensitive to 
the couplings. Actually, we introduced a deformation breaking the integrablity, and computed how the entanglement entropy of the level 1 states 
changes with the time $t$. For general couplings for which interactions are nonlocal, the entanglement entropy initially grows as $t^2$, but saturates 
at some value soon after and keeps oscillating. On the other hand, for the choice (\ref{alpha_expdecay}), the entanglement entropy keeps 
growing as $t^2$, and never reaches the point that is saturated in the nonlocal case. 
We can see that the exponential decreasing couplings crucially slow down the spreading of the entanglement.  
We are also considering to measure transport properties by computing connected two point correlation functions. 

The $SL(2,\,{\bf R})$ conformal symmetry plays a crucial role to construct the Hamiltonian (\ref{HN}) and thus to make the energy eigenstates 
independent of the couplings. Investigating this model from the viewpoint of AdS/CFT correspondence~\cite{chamon} will also be intriguing.

\begin{acknowledgement}
We would like to thank Catherine Meusburger for discussing the construction of integrable models by using coproducts. 
F.~S. would also like to thank the organizers and Vladimir Dobrev for invitation to the workshop. 
\end{acknowledgement}
%



\begin{thebibliography}{99.}%
\bibitem{altman}
E.~Altman and R.~Vosk, 
Ann.\ Rev.\ Condensed Matter Phys.\  {\bf 6} (2015) 383-409.   

\bibitem{anderson}
  P.~W.~Anderson,
  Phys.\ Rev.\  {\bf 109} (1958) 1492-1505.

\bibitem{ballesteros}
A.~Ballesteros and O.~Ragnisco,
J.\ Phys.\ A: Math.\ Gen.\ {\bf 31} (1998) 3791-3813.
  
\bibitem{basko}
D.~M.~Basko, I.~L.~Aleiner, B.~L.~Altshuler,
Ann.\ Phys. (NY) {\bf 321} (2006) 1126-1205. 
 

\bibitem{brandenberger} 
 R.~H.~Brandenberger,
  arXiv:1407.4775 [math-ph].
  
\bibitem{chamon}
C.~Chamon, R.~Jackiw, S.~Y.~Pi and L.~Santos,
 Phys.\ Lett.\ B {\bf 701} (2011) 503-507.


\bibitem{dealfano} 
V.~de Alfaro, S.~Fubini and G.~Furlan,
  Nuovo Cim.\ A {\bf 34} (1976) 569-612.


\bibitem{deutsch} 
J.~M.~Deutsch, 
Phys.\ Rev.\ A {\bf 43} (1991) 2046-2049. 

\bibitem{imbre0}
J.~Z.~Imbre, 
Jour.\ Stat.\ Phys. {\bf 163} (2016) 998-1048. 

\bibitem{imbre}
J.~Z.~Imbre, V.~Ros and A.~Scardicchio, 
Ann.\ Phys. (Berlin) {\bf 529} (2017) 1600278.

\bibitem{musso}
F.~Musso and O.~Ragnisco, 
J.\ Phys.\ A: Math.\ Gen.\ {\bf 34} (2001) 2625-2635. 


\bibitem{huse} 
R.~Nandkishore and D.~A.~Huse,
  Ann.\ Rev.\ Condensed Matter Phys.\  {\bf 6} (2015) 15-38. 
 
\bibitem{rigol}
M.~Rigol, V.~Dunjko and M.~Olshanii,
Nature {\bf 452} (2008) 854-858. 
 
  
\bibitem{srednicki}
M.~Srednicki, 
Phys.~Rev.~E {\bf 50} (1994) 888-901.

\bibitem{tasaki} 
H.~Tasaki, 
Phys.~Rev.~Lett. {\bf 80} (1998) 1373-1376. 



\end{thebibliography}
\end{document}